\documentclass[aps,prl,showpacs,twocolumn,groupedaddress,floatfix]{revtex4-1}
\usepackage{amssymb,amsmath}
\usepackage{graphicx}
\usepackage{subfigure}
\usepackage[english]{babel}
\usepackage{float}
\usepackage{color}

\begin{document}

\title{Expansion of a wave-packet in lattices with disorder and nonlinearity}

\author{Uta Naether$^{1}$, Santiago Rojas-Rojas$^{1}$, Alejandro J. Mart\'inez$^{1}$, Simon St\"utzer$^{2}$, Andreas T\"{u}nnermann$^{2}$, Stefan Nolte$^{2}$, Mario I. Molina$^{1}$, Rodrigo A. Vicencio$^{1}$, and Alexander Szameit$^{2}$}

\affiliation{$^{1}$Departamento de F\'isica, MSI-Nucleus on Advanced Optics, and Center for Optics and Photonics (CEFOP), Facultad de Ciencias, Universidad de Chile, Santiago, Chile}
\affiliation{$^{2}$Institute of Applied Physics, Friedrich-Schiller-Universit\"at Jena, Max-Wien-Platz 1, 07743 Jena, Germany}
\begin{abstract}
We show, theoretically and experimentally, the counterintuitive result that an increase of disorder can result in an enhanced spreading of an initially localized excitation. Moreover, we find that adding a focusing nonlinearity facilitates the expansion of the wave-packet even further by increasing its effective size. We find a clear transition between
between the regions of enhanced spreading (weak disorder) and localization (strong localization) described by a ``diffusion peak.''

\end{abstract}

\pacs{42.25.Dd, 42.65.Tg, 72.15.Rn}

\maketitle

The simultaneous presence of disorder and nonlinearity in wave systems can give rise to a variety of rich and complex behavior~\cite{rep1}. For a disordered system, it is a widespread belief that an increase in disorder will facilitate localization, that is, the initial spreading of an evolving wave-packet eventually stops~\cite{Anderson, Deissler}. When - in addition to disorder - a weak nonlinearity is added to the system, it has been predicted that the nonlinearity will either promote or inhibit the wave-packet spreading, depending on the systems details and the relative strength of disorder and nonlinearity~\cite{disnonl}. Thus, a general picture of the effects of the interplay between disorder and nonlinearity, remains unclear. Disorder-induced localization is based on wave interference and, hence, it is a universal concept applicable to a variety of physical systems~\cite{Sheng}, such as the transport of acoustic waves, microwaves, spin waves and matter waves~\cite{Weaver, Dalichaouch, Bruinsma, AndersonBEC}. In this respect, optical waveguide lattices~\cite{rep2} have emerged as ideal systems in which the interplay of disorder and nonlinearity can be observed by means of simple table-top experiments. After a first experimental study of wave evolution in a weakly-disordered nonlinear 2D fiber array~\cite{Led2D}, a ground-breaking work~\cite{naturesegev} was published where the averaging over multiple individual realizations of disorder played a key role, resulting for the first time in the experimental demonstration of genuine Anderson-localization. In that work it was also shown that focusing nonlinearity indeed facilitates localization. Whereas this work was performed in a two-dimensional (2D) system, in a subsequent experiment~\cite{1Dexp} the impact of nonlinearity on localization in a one-dimensional (1D) system was experimentally analyzed,
culminating in the same result: focusing nonlinearity indeed enhances localization of propagating waves.


In this work we present theoretical and experimental evidence that the presence of weak disorder can promote the spreading of a wave-packet. Moreover, this effect is further enhanced by the addition of focusing nonlinearity. We focus on finite lattices, where the  system dimensions can be smaller than the localization length, and monitor the spreading of the wave-packet during the first stages of its evolution.

Longitudinal light propagation in weakly-coupled, nonlinear optical waveguide arrays can be  modeled by a set of normalized, discrete nonlinear Schr\"{o}dinger (DNLS) equations~\cite{rep1,rep2}:
\begin{equation}
-i\frac{d u_{{\bf n}}}{d z} =\epsilon_{{\bf n}} u_{{\bf n}}+ \sum_{{\bf m}} V_{{\bf n},{\bf m}}u_{{\bf m}} + |u_{{\bf n}}|^2 u_{{\bf n}} .
\label{model}
\end{equation}
Here, $u_{{\bf n}}$ is the amplitude of the waveguide mode in the {\it {\bf n}}-th waveguide. The coordinate ${\bf n}$ depends on the dimension and lattice type [for example, ${\bf n}=(k,l)$ in a 2D square lattice]. The quantity $\epsilon_{{\bf n}}$ is the propagation constant (i.e., ``site energy'' in the quantum mechanical context) of the {\it ${\bf n}$}-th guide. The hopping between adjacent lattice sites ${\bf n}$ and ${\bf m}$ is described by the coupling constant $V_{{\bf n},{\bf m}}$. In our model we impose disorder on both, the propagation constant [$\epsilon_{{\bf n}} \in \{-W_{\epsilon}/2,W_{\epsilon}/2\}$] and the coupling constants [$V_{{\bf n},{\bf m}} \in 1+\{-W_{c}/2,W_{c}/2\}$]; the disorder is therefore characterized by the disorder strengths $W_{\epsilon}$ and $W_{c}$, respectively. Note that we limit $W_{c}$ to the interval $\{0,2\}$, to insure that the inter-site coupling is always positive.
\begin{figure}[t]
\begin{center}
\includegraphics[width=0.45\textwidth]{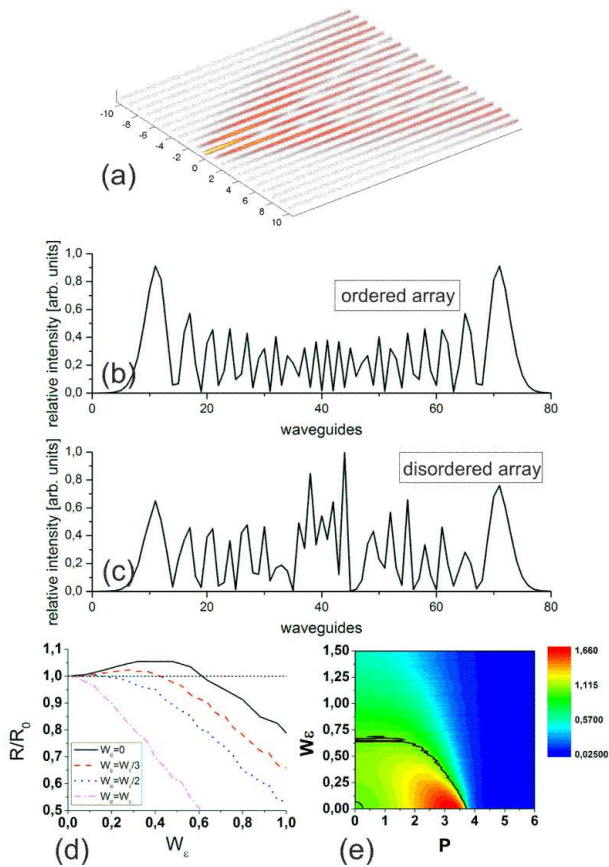}
\caption{(color Online). (a) Sketch of the light propagation in a 1D waveguide array, when only one waveguide is excited. (b)-(c) Simulated intensity distribution in the lattice sites for ordered (b) and on-site disorder level $W_{\epsilon}=0.72$ (d) after propagation through the waveguide array when only the central waveguide is excited. (d) Simulated averaged PR $R$ vs. on-site disorder for different levels of inter-site disorder after linear (low power) propagation. (e) Simulated averaged PR  vs. Power $P$ and onsite-disorder $W_{\epsilon}$ after nonlinear propagation (thick contour denotes $R=R_0$). In all cases, the PR is normalized to $R_0\equiv R(W_{\epsilon}=0)$.}\label{fig1}
\end{center}
\end{figure}
One of the most useful measures used for the description of localization phenomena in finite systems is the participation ratio
(PR)~\cite{sciencesegev} which gives a rough estimate of the number of sites that are occupied by the wave-packet, i.e., where the light has a significant amplitude at the output facet of the sample. As pointed out
in~\cite{sciencesegev}, in order to get meaningful data for finite lattices, one has to also average over different realizations of disorder, and thus, work with an averaged PR: $R=\left \langle \sum_{{\bf n}=1}^{N} |u_{{\bf n}}|^{4}/P^2\right \rangle^{-1}$ \cite{PR}, where $P$ is the total optical power, and $N$ is the total number of waveguides.

We start our analysis by carrying out extended numerical simulations in 1D lattices. We integrate Eq. \eqref{model}, for many random realizations, and perform an averaging process for the relevant quantities. In the 1D case, we use $N=81$ waveguides with an initially localized excitation:
$u_n(0)=\sqrt{P}\delta_{n,n_c}$, where $n_c$ corresponds to the input position. The participation ratio was taken at the propagation distance $z_{max}=20$ (in normalized units). In order to simulate linear propagation, the total optical power $P$ is chosen very small ($<0.01$).
Figure~\ref{fig1}(a) shows light propagation in an ordered waveguide array. In order to reproduce Anderson's results \cite{Anderson}, we show in Fig.~\ref{fig1}(b)-(c)  output distributions for the ordered lattice and one level of pure on-site disorder, showing a ballistic spreading for the ordered ($W_{\epsilon}=0$) case [Fig.~\ref{fig1}(b)]. The diffraction pattern exhibits distinct side lobes and small amplitudes around the initially excited central site. When weak disorder is introduced [$W_{\epsilon}=0.72$, Fig.~\ref{fig1}(c)], the power contained in these lobes is redistributed to the waveguides close to the center. By increasing the disorder further, a localization around the excited site is observed, with exponentially decreasing amplitudes on both sides. However, an important aspect here  is the occupation of the individual lattice sites, that defines the PR, i.e. the effective width of the diffraction pattern. Our simulations reveal something very surprising: As the (weak) disorder strength is increased, the average PR increases  [black solid line in Fig.~\ref{fig1}(d)]. Upon further increment of the disorder, the spatial profile reduces its expansion and the PR decreases. The figure shows a clear maximum expansion which separates regimes of expansion and localization; i.e, a ``diffusion-peak''. We note in passing that, even though this phenomenon was encountered before~\cite{disPRE}, it was neither discussed in depth nor explained.

We also investigate the role of mixed disorder by increasing the amount of inter-site disorder $W_{c}$. The results for $W_c=W_{\epsilon}/3,W_{\epsilon}/2$, and $W_{\epsilon}$, are shown in Fig.~\ref{fig1}(d) as the red dashed line, the blue dotted line, and the pink dash-dotted line, respectively. From our simulations it is evident that, as the inter-site disorder increases, the initial growth of the PR is reduced, and eventually vanishes (i.e. $R/R_0<1$ for all disorder levels) for sufficiently high inter-site disorder.

Next, we analyze the impact of nonlinearity, i.e. the propagation of high-power optical  beams.  In general, one would expect that a focusing nonlinear term in Eq. \eqref{model} would facilitate the self-focusing of the excitation around the initially excited site~\cite{1DSoliton}. We find that, in contrast, in the weak-disorder regime the increase of the PR is enhanced in the presence of a small amount of nonlinearity.  Figure \ref{fig1}(e) shows the average participation ratio of the profile after propagating a certain distance in the presence of on-site disorder $W_{\epsilon}$ and power (nonlinearity) $P$ in a 1D lattice. Interestingly, when the disorder is switched off and only nonlinearity remains, the picture is similar: the PR increases up to some maximum value due to redistribution of the power in the waveguides and then drops as the wave-packet localizes~\cite{ST1D}. Therefore, below a critical value ($P<4$), nonlinearity can facilitate the delocalization process.

\begin{figure}[t]
\begin{center}
\includegraphics[width=0.45\textwidth]{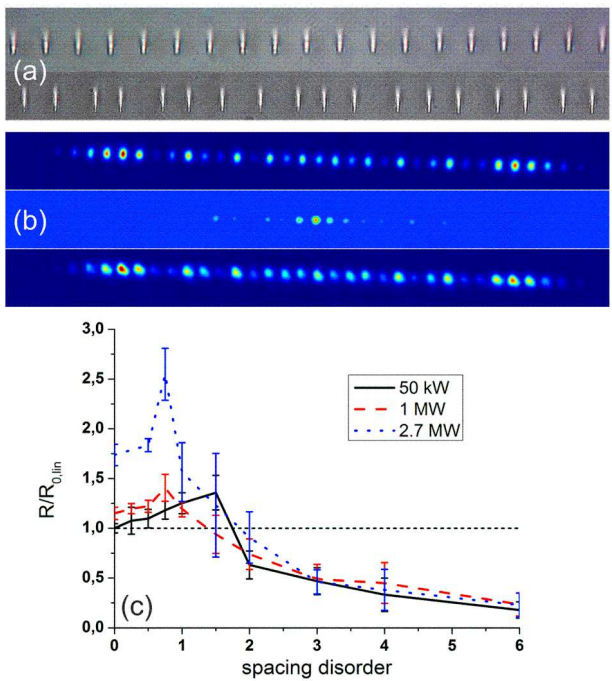}
\caption{(color Online). (a) Microscope images of an ordered and disordered 1D array. (b) Experimental intensity output patterns. Upper row: ordered array; middle row: single realization disordered array; bottom row: averaged disordered output pattern. (c) Experimental PR $R$ versus disorder strengths for different levels of input power. }\label{fig2}
\end{center}
\end{figure}

These theoretical predictions are confirmed by our experiments, whose results are summarized in Fig.~\ref{fig2}. The experiments are carried out in waveguide arrays fabricated by the direct-writing laser technology~\cite{arrays} in polished bulk fused silica wafers. The dimensions of each guide are ~$4\times12$ $\mu \mbox{m}^2$ with a refractive index increase of $\approx 5\times10^{-4}$ and a propagation length of $100$ $\mbox{mm}$. For the analysis of 1D samples, we fabricated several arrays with $N=81$ sites each: One without disorder, and nine with varying degrees of disorder (and $30$ realizations for each degree of disorder). Disorder was created by varying the spacing between the guide centers: $d=(23\pm\delta_d) [\mu \mbox{m}]$, $\delta_d=(0,0.25,0.5,0.75,1,1.5,2,3,4,6)$. The difference in overlap of the individual waveguide modes additionally creates a statistic detuning of the guides and, therefore, creates an additional on-site disorder.
%
At the input facet, light at $\lambda=800$ $\mbox{nm}$ from a Ti:Sapphire laser was launched into a single waveguide in the center of the array using a standard microscope objective. At the end facet of the sample, the intensity patterns were recorded with a CCD camera. Whereas the linear regime ($P_{lin}$) was measured by using the continuous wave-mode of the laser (i.e., the laser power was only a few $\mbox{mW}$), the nonlinear regime was studied in the pulsed regime with pulse peak power of $P_{nl1}\simeq1$MW and $P_{nl2}\simeq2.7$MW. The participation number $R$ and its standard deviation were
computed for all the different settings, and normalized to the value of the linear single-site excitation in an ordered array ($R_{0,lin}$). A microscope image of the front facet of an ordered and a disordered 1D lattice is shown in Fig.~\ref{fig2}(a). Fig.~\ref{fig2}(b) shows different propagation patterns of an ordered array (upper row), a single disordered realization (middle row), and an averaged output profile for the same disorder level (lower row). From each realization, the PR was computed and then averaged. As it can be clearly seen [see Fig.~\ref{fig2}(c)], in the linear regime, we observe the delocalization tendency with the increase of weak disorder and a diffusion-peak at a disorder level of $\delta_d=1.5$ $\mu \mbox{m}$ spacing variation between the guides. When nonlinearity comes into play, the diffusion-peak shifts towards smaller disorder levels, and its height increases significantly. However, the delocalization is only observed for sufficiently small nonlinearities. If the nonlinearity is too high, the delocalization effect vanishes as the self-focusing is sufficiently strong to inhibit any diffusion/expansion process.

\begin{figure}[t]
\begin{center}
\includegraphics[width=0.45\textwidth]{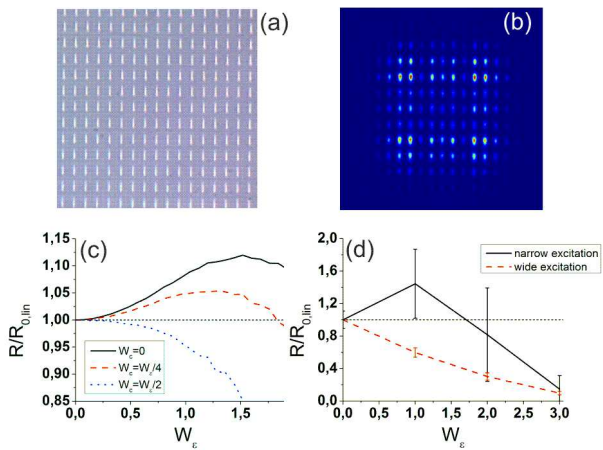}
\caption{(color Online). (a) Microscope image of the fabricated ordered square lattice. (b) Corresponding averaged output intensity profile. (c) Numerical simulation of the PR for different mixed disorder widths: $W_c=0$ (black solid line), $W_c=W_{\epsilon}/4$ (red dashed line), $W_c=W_{\epsilon}/2$ (blue dotted line). (d) Experimental results for single-site (black solid line) and gaussian (red dashed line) input excitations, averaged over 25 realizations. All data is normalized to the ordered array $R_{0}$. }\label{fig3}
\end{center}
\end{figure}

In 2D waveguide arrays, one of the simplest structures is the square lattice where each site is coupled to four nearest-neighbors only.
We fabricated lattices with $21\times 21$ sites with three different levels of disorder and one ordered array. The mean distances $d_{hor}=(17\pm\delta_h)[\mu m]$ with $\delta_{h}=(0,2,4,6)$. Although the vertical distance was kept fixed at $d_v=23[\mu m]$, this setting is equivalent to a 2D fully disordered lattice~\cite{simon}. Figure ~\ref{fig3}(a) shows a microscope image of an ordered square waveguide lattice, while Fig.~\ref{fig3}(b) shows its experimental output intensity pattern for a single-guide excitation. It exhibits four distinct side lobes. Figure~\ref{fig3}(c) shows numerical simulation results for the disordered lattice case. They are qualitatively similar to the ones obtained in 1D lattices, with the height and position of the diffusion peak decreasing as mixed disorder increases. Our experimental results, summarized in Fig.~\ref{fig3}(d), validate this scenario: For small inter-site disorder the PR significantly grows, and for further increase of the disorder the PR drops, resulting in localization of the wave-packet (black solid line). Importantly, distinct side lobes in the diffraction pattern only occur when a single waveguide is excited~\cite{rep1,rep2}. Numerical simulations show that a wide Gaussian beam creates a propagation pattern with a single lobe at the center, whose height (width) decreases (increases) continuously during propagation. Its PR decreases monotonically with increasing disorder. This was experimentally verified by using a Gaussian input beam that covered  approximately 9 sites. We clearly see that there is no delocalization enhancement for such initial condition [Fig.~\ref{fig3}(d), red dashed line]. As for pure inter-site disorder, no diffusion-peak is observed in our simulations, we therefore conclude that the variation of the inter-site spacing results in a mixed disorder in the fabricated samples.

Most explanations concerning diffusion processes in disordered lattices resort to band structure changes. Such is the case of disordered quasicrystal structures~\cite{sciencesegev}. However, in this case the reduction of the pseudo-gaps due to disorder does not necessarily imply that a particular excitation located at any input position will be able to excite more states.
For weak disorder it is not possible to claim that, by considering a single-site excitation, a particular state will get excited; this may be only possible for strong disorder when all states are very localized. Therefore, in our view it is not useful to discuss a precise excitation of states, as this would be certainly a matter of probability~\cite{flachArXiv} and it will not correspond to a representative case. Instead, we rely on simple dynamical arguments. A wave-packet initially localized in a single waveguide can be decomposed into a superposition of plane waves, each one with a different transverse velocity. In the absence of disorder, during propagation this wave-packet will form the typical side lobes of the diffraction pattern~\cite{rep1,rep2} where the waves possess a high transverse velocity, and most of the propagating power is concentrated in these lobes. For weak disorder, one can assume that launching light into a single waveguide excites a number of different linear modes with different transverse velocities. The main lobes still propagate without strong distortions as those plane waves possess sufficient kinetic energy to overcome the disordered potential wells and to move across the lattice. However, some waves with less kinetic energy will get trapped in the impurity regions of the particular disorder distribution. As a consequence, the wave will expand more homogeneously, localizing energy in the input region (small velocity waves), in the lobes (high velocity waves), and in between
(intermediate velocities). When the disorder is further increased, the waves in the side lobes will also get trapped as their energy becomes comparable to the deeper disordered potential wells. Therefore, light reduces its expansion and the localization starts to show its signature [see Fig.~\ref{fig1} and Refs.~\cite{Led2D,naturesegev,1Dexp}]. In this light, the PR will increase at small disorder strength and decrease only at sufficiently high disorder. It is important to clarify that this process does not have implications in the \textit{edge-to-edge-diameter} of the wave-packet; it rather implies a denser distribution of the light and therefore a higher PR.
The side lobes remain the dominating feature of the propagation pattern even in the far field~\cite{Longhi}. Also,  the effect is not connected to the initial diffusive spreading at short propagation distances~\cite{Izrailev}.

In the absence of disorder, the side lobe-dominated energy distribution mechanism is also found in higher dimensional lattices: 2D square and honeycomb, and 3D lattices - all of these structures will therefore exhibit the initial increase of the PR for small disorder and, therefore, a diffusion-peak. An exception is the triangular (hexagonal) lattice which, due to its high coordination number (6), possesses an unusual discrete diffraction pattern with no distinct lobes, resembling the diffraction pattern for a continuous medium. Thus, our numerical simulation of the triangular lattice finds no enhancement of the PR, even at weak disorder, in agreement with the experimental findings in~\cite{naturesegev}.

For a disorder level in the vicinity of the diffusion-peak, the addition of a weak nonlinearity will increase the (random) refractive index at each site, in an amount proportional to the light intensity on the site. This causes the high-intensity side lobes to be affected the most, whereas the `slow' waves in the center are not affected much. However, the latter are already localized, as their energy content is insufficient to overcome the disordered potential wells. Thus, the nonlinear deepening of the random potential wells, renormalizes the disorder strength of the individual lattice sites. This, in turn, causes an increment of the participation ratio which occurs now at smaller values of disorder strength than in the linear case, in agreement with our experimental and numerical results.

In summary, we have shown both, theoretically and experimentally that the presence of weak disorder can lead to an enhancement of the spreading of an initially localized optical beam in 1D and 2D waveguide arrays. Moreover, the addition of focusing nonlinearity facilitates the expansion even further. The regions separating enhanced expansion (weak disorder) and localization (strong disorder) can be clearly separated by a ``diffusion peak'', whose general behavior can be explained by dynamical arguments.


The authors wish to thank Markus Gr\"afe for useful discussions. This work was supported in part by FONDECYT Grants 1110142, 1080374, CONICYT fellowships, Programa ICM P10-030-F, the Programa de Financiamiento Basal de CONICYT (FB0824/2008), the Deutsche Forschungsgemeinschaft (grant NO 462/6-1), and the German Ministry of Education and Research (Center for Innovation Competence program, grant 03Z1HN31).

\end{document}